\documentclass[10pt]{scrartcl}
\usepackage[T1]{fontenc}
\usepackage[latin1]{inputenc}

\makeatletter

\providecommand{\LyX}{L\kern-.1667em\lower.25em\hbox{Y}\kern-.125emX\@}
\newcommand{\noun}[1]{\textsc{#1}}

\usepackage{amsmath,amsfonts} 

\renewcommand{\hat}[1]{\mathsf{#1}}

\makeatother

\begin{document}

\title{Finite-time Lyapunov exponents for products of random transformations
\footnote{\emph{J. Stat. Phys.}, {\bf 112} (2003) 193--218}}

\author{Andrea Gamba~\thanks{
Dipartimento di Matematica, Politecnico, 10129 Torino, Italia. E-mail: \texttt{gamba@polito.it}.
}}

\maketitle
\begin{abstract}
It is shown how continuous products of random transformations constrained by a generic
group structure can be studied by using Iwasawa's decomposition into ``angular'',
``diagonal'' and ``shear'' degrees of freedom. In the  case of a Gaussian
process a set of variables, adapted to the Iwasawa decomposition and still having
a Gaussian distribution, is introduced and used to compute the statistics of
the finite-time Lyapunov spectrum of the process. The variables also allow to
show the exponential freezing of the ``shear'' degrees of freedom, which contain
information about the Lyapunov eigenvectors. 
\end{abstract}
\setlength{\textwidth}{27pc} \setlength{\textheight}{43pc}

\section{Introduction}

Lyapunov exponents represent the natural way to measure the rapidity of divergence
of a set of infinitesimally close points carried along by a chaotic or random
flow \cite{aw86,cpv93,ak87}. The simplest example is given when a smooth \( N \)-dimensional
dynamical system of the form 
\[
\dot{\mathbf{x}}=\mathbf{F}\left( \mathbf{x}\right) \]
 is linearized around a trajectory \( \mathbf{x}\left( t\right)  \) giving
the linear equation \( \dot{\mathbf{y}}=X\left( \mathbf{x}\right) \mathbf{y} \)
for the separation \( \mathbf{y}\left( t\right)  \) between two infinitesimally
close orbits. The generator of the tangent dynamics, \( X_{ij}=\partial _{i}F/\partial x_{j} \),
is known as the stability matrix of the system. The evolution of the initial
separation \( \mathbf{y}\left( 0\right)  \) is formally given by the expression
\[
\mathbf{y}\left( t\right) =g\left( t\right) \mathbf{y}\left( 0\right) \]
where \( g\left( t\right)  \) is the time-ordered product
\begin{equation}
\label{texp}
g\left( t\right) =\mathrm{T}\exp \left[ \int _{0}^{t}X\left( t'\right) \, \mathrm{d}t'\right] 
\end{equation}
The \( N \) time-dependent Lyapunov exponents \( \lambda _{1}\left( t\right) \ge \cdots \ge \lambda _{N}\left( t\right)  \)
are defined as the eigenvalues of the symmetric matrix \( \left( 2t\right) ^{-1}\log \left[ g^{\dagger }\left( t\right) g\left( t\right) \right]  \),
and measure the rate of growth of infinitesimal \( k \)-dimensional volumes
(\( k=1,\ldots ,N \)). Oseledec \cite{oseledec68} proved under very general
conditions the existence of the \( t\rightarrow \infty  \) limits of the \( \lambda _{k}\left( t\right)  \)
and of the eigenvectors \( \mathbf{f}_{k}\left( t\right)  \) of \( g^{\dagger }\left( t\right) g\left( t\right)  \). 

A common \emph{ansatz} for strongly chaotic systems assumes that the stability
matrices \( X\left( t\right)  \) decorrelate in a characteristic time, shorter
with respect to other time scales of the system, and can therefore be treated
as a succession of independent linear transformations \cite{cpv93}. Then the
separation \( \mathbf{y}\left( t\right)  \) evolves according to the stochastic
equation

\begin{equation}
\label{lang}
\dot{\mathbf{y}}=X\left( t\right) \mathbf{y}
\end{equation}
 After carefully specifying the statistics of the random process \( X\left( t\right)  \),
and choosing in particular one of the standard (It\^o or Stratonovich) regularization
for (\ref{lang}), the averages of functionals \( f\left[ \mathbf{y}\left( t\right) \right]  \)
over the random realizations \( \mathbf{y}\left( t\right)  \) can be written
in the form of path integrals \cite{zinn96}:

\begin{equation}
\label{ave}
\left\langle f\left[ \mathbf{y}\left( t\right) \right] \right\rangle =\int \mathcal{D}X\, f\left[ \mathbf{y}\left( t\right) \right] \, \mathrm{e}^{-S\left[ X\right] }
\end{equation}
The problem of computing averages of the form (\ref{ave}) is quite general,
and appears both in the context of quantum and classical systems. 

In the theory of quasi one-dimensional disordered wires \cite{beenakker97}
the study of low-temperature conductance fluctuations in small metallic samples
is reduced to the computation of the statistics of \( g^{\dagger }\left( L\right) g\left( L\right)  \),
where \( g\left( L\right)  \) is the \( 2n\times 2n \) transfer matrix for
a wire of length \( L \) and \( n \) is the number of transverse modes at
the Fermi level \cite{dorokhov82}. The transfer matrix \( g\left( L\right)  \)
can be naturally written in the form (\ref{texp}) as an ordered product of
transfer matrices for infinitesimal portions of wire. Here the role of time
is played by the length \( L \) of the sample, and the role of the Lyapunov
exponents by (inverse) localization lengths. Ensemble averaging implies the
computation of path integrals of the form (\ref{ave}), where the infinitesimal
transfer matrices \( X \) are constrained by symmetry considerations. In the
presence of time reversal symmetry the \( X \) matrices belong to the \( \hat{s}\hat{p}\left( n,\mathbf{R}\right)  \)
algebra, if time reversal symmetry is broken they belong to \( \hat{s}\hat{u}\left( n,n\right)  \),
and in the presence of spin-orbit scattering to \( \hat{s}\hat{o}^{*}\left( 4n\right)  \)
\cite{huffman90}. 

In the theory of passive scalar transport in \( n \)-dimensional space \cite{fgv01}
the computation of many statistical quantities, such as the exact p.d.f.'s of
the scalar \cite{cfkl95a,cgk94,gk96,bgk98} and of its gradients \cite{cfk98,gk99},
can be reduced to the form (\ref{ave}), where \( X \) is a matrix of velocity
gradients of the carrying fluid. The matrices \( X \) are the infinitesimal
generators of the dynamics of an incompressible fluid element and belong to
the \( \hat{s}\hat{l}\left( n,\mathbf{R}\right)  \) algebra.

In the study of deterministic chaos, products of random matrices have been used
to mimic the chaoticity of dynamical systems \cite{cpv93}. In the case of \( n \)-dimensional
Hamiltonian systems one has products of random symplectic maps \cite{benettin84,pv86}.
This formally leads to expressions of the form (\ref{ave}) where the matrices
\( X \) belong to the \( \hat{s}\hat{p}\left( n,\mathbf{R}\right)  \) algebra.
Similar models have been recently used to study the hydrodynamic modes in the
Lyapunov spectrum of extended dynamical systems \cite{eg00}. 

In many of these cases analytic results were obtained. Such results were made
possible by the presence of three main ingredients: \emph{i)} Markovianity\footnote{%
Although results were obtained also with time-correlated statistics, see \emph{e.g.}
Refs.~\cite{cfkl95a,fkl98}.
}, \emph{ii)} Gaussianity, and \emph{iii)} the presence of a group structure
in the set of transformations and of a statistics compatible with it (\emph{i.e.},
a statistics that can be expressed in terms of the intrinsic geometry of the
transformation group). 

While the physical support for the first two assumptions is in some cases questionable,
the existence of a group structure lays on stronger foundations, since it directly
reflects the symmetries of the underlying physical problem.

In the case of quasi one-dimensional wires, results were obtained studying the
Fokker-Planck equation for the transmission eigenvalues as a diffusion equation
over Riemannian manifolds \cite{br93a, caselle95}. A similar approach has been
also applied to the case of passive scalar transport \cite{bgk98}. On the other
hand, functional integral representations of the form (\ref{ave}) for the probability
densities can be exploited directly \cite{kolokolov93, cgk94a, cfkl95a,cgk94,cfk98,gk99,cfkv99}
and can provide an intuitive description of the statistics of the Lagrangian
trajectories described by Eq.~(\ref{lang}). Path integral representations also
provide the ground for several approximations, like instanton calculus and perturbative
expansions. Actually, the two approaches are complementary and one should switch
from one to the other depending on convenience. However, the Lagrangian theory
has been worked out mainly on specific examples, often relying on particular
features of the given algebras \cite{gk96,gk99}, and not covering the most
general case. The purpose of this work is to present this theory in a sufficiently
general setting. It turns out that a formal exposition not only allows to cover
many different cases in a unified way, but also unveils the simple geometric
meaning of some formul\ae\ and presents a very simple and appealing scenario,
which can hopefully serve as a basis for further investigations. 

The starting point of the Lagrangian theory is a convenient decomposition of
the time-ordered exponential (\ref{texp}). Let us remind here some facts about
the standard method \cite{bggs80} for the numerical computations of Lyapunov
exponents. An orthonormal set of \( k \) vectors is evolved for a time \( t \)
and a Gram-Schmidt orthogonalization procedure is applied to the evolved vectors
in order to obtain a new orthonormal set. The procedure is iterated, and by
performing an average of the logarithmic \( k \)-volume increase over many
steps one obtains the approximated sum of the first \( k \) Lyapunov exponents.
When the evolution operator \( g\left( t\right)  \) belongs to the \( \mathrm{SL}\left( n,\mathbf{R}\right)  \)
group it is easy to see \cite{helgason78} that the Gram-Schmidt orthogonalization
procedure is equivalent to the decomposition of the evolution operator in the
product or three consecutive transformations, corresponding to a shear, a dilation,
and a rotation: the rotation carries the old orthonormal set in the new one,
the dilation is responsible for the volume increase, and the shear part inherits
from the Gram-Schmidt procedure a triangular form. From an algebraic point of
view, the group element \( g\left( t\right)  \) is decomposed into the product
of a compact, Abelian and nilpotent part. Iwasawa \cite{iwasawa49} showed that
a similar decomposition can be defined for a large class of groups and is a
smooth mapping (see Ref. \cite{helgason78} and App. \ref{liecartan}). 

The fact that emerged from a series of works \cite{kolokolov86,kolokolov90,cgk94,gk96,gk99}
is that starting from normally distributed, centered random variables \( X \)
it is possible in certain cases to introduce new variables, adapted to the Iwasawa
decomposition\footnote{%
Or to the Gauss decomposition, which was considered in Refs. \cite{kolokolov86,cgk94}
respectively for the \( \hat{s}\hat{u}\left( 2\right)  \) and \( \hat{s}\hat{u}\left( n\right)  \)
cases.
}, \emph{which are still normally distributed}. Here it is shown that the new
variables can be introduced in the quite general case where the random variables
\( X \) belong to the algebra of a real semisimple Lie group, and their covariance
has a form compatible with the group structure (\emph{cf.} Eq. (\ref{cova})).
It turns out that the definition of the new variables (the \( \{A,M\} \) variables
defined by \emph{}Eqs. (\ref{IWA},\ref{rotated})) is most easily given in
terms of the group structure of the set of transformations, and in particular
of the ``adjoint operator'' \( \mathrm{Ad}\left( g\right) X=gXg^{-1} \). 

The fact that the new variables are normally distributed is non trivial, since
the Iwasawa transformation is nonlinear, and it is of course quite relevant
in practical calculations. In particular, it allows the rapid computation of
the Lyapunov spectrum, which is encoded in the statistics of the Abelian part
of the Iwasawa decomposition. Although the original variables \( X \) are centered,
the Abelian part acquires a positive mean value thanks to the non-trivial Jacobian
factor associated with the variable transformation (\emph{cf.} (\ref{locfac})).
This mean value is precisely the finite-time Lyapunov spectrum, which can be
expressed explicitly in terms of the structure constants of the algebra of the
\( X \) variables (\emph{cf.} Eqs. (\ref{media},\ref{defcov})).

The new variables however contain more information than just the statistics
of Lyapunov exponents. For instance, Eqs. (\ref{esprexp},\ref{media},\ref{gred2})
prove in a quite general setting the freezing of the ``shear'' (nilpotent)
degrees of freedom which was used in Refs. \cite{cfk98,gk99} to compute the
statistics of passive scalar dissipation. The nilpotent degrees of freedom actually
encode information about the Lyapunov eigenvectors \( \mathbf{f}_{k}\left( t\right)  \).
For instance, in the \( \mathrm{SL}\left( n,\mathbf{R}\right)  \) case it is
easy to see that the \( \mathbf{f}_{k}\left( t\right)  \) can be obtained (up
to exponentially small terms) by Gram-Schmidt orthogonalization of the row vectors
of the (nilpotent) shear matrix. The exponential freezing of the shear part
therefore implies the exponentially rapid convergence of the eigenvectors \( \mathbf{f}_{k}\left( t\right)  \)
to their limits for \( t\rightarrow \infty  \), thus providing a particularly
clear realization of Oseledec theorem.

The work is organized as follows. In Sec. \ref{due}, I define the random process
\( X\left( t\right)  \) by assigning its covariance in terms of intrinsic geometric
features of the corresponding algebra. This is equivalent to assigning a Feynman-Kac
measure on the space of paths \( t\rightarrow X\left( t\right)  \). The multiplicative
process \( g\left( t\right)  \) is then defined as a time-ordered exponential
through the appropriate (Stratonovich) regularization. In Secs. \ref{tre},\ref{quattro},\ref{cinque}
the new variables are defined through the following steps: \emph{i)} performing
the Iwasawa decomposition of the process \( g\left( t\right)  \); \emph{ii)}
introducing intermediate variables, adapted to the Iwasawa decomposition; \emph{iii)}
rewriting the Gibbs weight \( \exp \left( -S\right)  \) in terms of the new
variables; \emph{iv)} computing the Jacobian factor of the variable transformation,
\emph{i.e.}, the transformation rule for the ``volume form'' \( \mathcal{D}X \);
\emph{v)} performing a Lie algebra transformation that leads to new, normally
distributed variables. Although the intermediate steps are technical, the final
results are simple and are summarized by formul\ae\ (\ref{gred}-\ref{covarianza}).
In Sec. \ref{sei} the new variables are used to compute the statistics of the
finite time Lyapunov spectrum of a generic linear representation of the process
\( g\left( t\right)  \). In Sec. \ref{sette} the cases of the \( \hat{s}\hat{l}\left( n,\mathbf{R}\right)  \)
and \( \hat{s}\hat{p}\left( n,\mathbf{R}\right)  \) algebras are worked out
in detail. A brief summary of results of Lie algebra theory are given in the
Appendix.

\section{Random walks on Lie groups\label{due}}

The computational scheme outlined in the Introduction can be implemented in
the quite general case where the set of transformations \( g\left( t\right)  \)
form a real semisimple Lie group \( G \) (this comprises the case of complex
groups, as soon as they are seen as real groups endowed with a complex structure).
The infinitesimal generators \( X\left( t\right)  \) then lie in the corresponding
algebra \( \hat{g} \). Each real semisimple Lie algebra is characterized by
two natural structures: a canonical quadratic form \( B\left( X,Y\right)  \)
known as the Killing form, and an involution \( \theta  \) such that \( B\left( X,\theta Y\right)  \)
becomes strictly negative definite (see Appendix \ref{liecartan}). For instance,
in the case \( \hat{g}=\hat{s}\hat{l}\left( n,\mathbf{R}\right)  \) (the algebra
of real \( n\times n \) real matrices with null trace) \emph{}one has \emph{\( B\left( X,Y\right) =2n\mathrm{Tr}XY \),}
\( \theta X=-X^{\mathrm{tr}} \) and the effect of \( \theta  \) is that of
changing the sign of the symmetric part of \( X \). \emph{}

In order to define a Gaussian probability density on \( \hat{g} \) one needs
a positive definite quadratic form. I will consider here the general linear
combination
\begin{equation}
\label{metric}
F\left( X,Y\right) =\mu B\left( X,Y\right) +\nu B\left( X,\theta Y\right) 
\end{equation}
 where \( \mu  \), \( \nu  \) are real parameters such that \( F \) be positive
definite (see (\ref{positivo}) below).

Let \( X\left( t\right)  \) be the Gaussian process on \( \left[ 0,T\right]  \)
with values in \( \hat{g} \), zero mean and covariance 
\begin{equation}
\label{cova}
\left\langle X\left( s\right) \otimes X\left( t\right) \right\rangle =\delta \left( t-s\right) \cdot C
\end{equation}
 where \( C=F^{-1} \), \emph{i.e.} the random process defined by the Feynman-Kac
measure

\begin{equation}
\label{meas}
\mathrm{d}\mu \left( X\right) =K\, \exp \left[ -\frac{1}{2}\int _{0}^{T}F\left( X\left( t\right) ,X\left( t\right) \right) \mathrm{d}t\right] \prod _{t=0}^{T}\mathrm{d}X\left( t\right) 
\end{equation}
 where \( \mathrm{d}X \) is a Euclidean measure on \( \hat{g} \) and \( K \)
is an (infinite) normalization constant. The continuous process \( X\left( t\right)  \)
can be seen \cite{fh65,gy60} as the limit of the discrete process \( X_{j} \),
where \( X_{j}=X\left( t_{j}\right)  \), \( t_{j}=j\epsilon  \), \( \epsilon =T/N \),
defined by the probability measure 
\begin{equation}
\label{discrmeas}
\mathrm{d}\mu _{N}\left( X\right) =K_{N}\, \exp \left[ -\frac{1}{2}\sum _{j=1}^{N}\epsilon \, F\left( X_{j},X_{j}\right) \right] \prod _{j=1}^{N}\mathrm{d}X_{j}
\end{equation}
 as \( N\rightarrow \infty  \) .

Let \( g\left( t\right)  \) be the time-ordered exponential defined by 
\[
g\left( t\right) =\mathrm{T}\exp \left( \int _{0}^{t}X\left( \tau \right) \mathrm{d}\tau \right) \equiv \lim _{N\rightarrow \infty }g_{N}\]
 where \( g_{j} \) is defined recursively by 
\begin{equation}
\label{rec}
g_{j}=\exp \left( \epsilon X_{j}\right) g_{j-1},\qquad g_{0}=e\: (\textrm{the identity})
\end{equation}
This discretization prescription is compatible with the group structure, and
is of Stratonovich type. As a matter of fact, if \( G \) is a linear group
(and, in general, on the universal enveloping algebra of \( \hat{g} \)) one
can compute
\begin{eqnarray*}
 &  & \frac{1}{2\epsilon }\left( g_{j}-g_{j-1}\right) \left( g_{j}^{-1}+g_{j-1}^{-1}\right) \\
 & = & \frac{1}{2\epsilon }\left( g_{j}g_{j-1}^{-1}-g_{j-1}g_{j}^{-1}\right) \\
 & = & \frac{1}{2\epsilon }\left[ \exp \left( \epsilon X_{j}\right) -\exp \left( -\epsilon X_{j}\right) \right] \\
 & = & X_{j}+O\left( \epsilon ^{2}X_{j}^{3}\right) 
\end{eqnarray*}
 obtaining 
\begin{equation}
\label{midpoint}
X_{j}=\frac{g_{j}-g_{j-1}}{\epsilon }\frac{g_{j}^{-1}+g^{-1}_{j-1}}{2}+O\left( \epsilon ^{2}X_{j}^{3}\right) 
\end{equation}
 where \( O\left( \epsilon ^{2}X_{j}^{3}\right)  \) denotes a series of terms
having order at least 2 in \( \epsilon  \) and at least 3 in \( X_{j} \).
Such terms can be neglected in averages when \( N\rightarrow \infty  \), as
usual in the theory of Feynman-Kac integrals \cite{fh65}, since \( \left\langle X_{j}\otimes X_{l}\right\rangle =\epsilon ^{-1}\delta _{jl}\cdot C\sim 1/\epsilon  \).
In other words, in the limit \( N\rightarrow \infty  \) one can consider 
\[
X\left( t\right) =\dot{g}\left( t\right) g^{-1}\left( t\right) \]
 understanding that the midpoint regularization (\ref{midpoint}) has to be
used throughout.

\section{Decomposition of the trajectories\label{tre}}

Following the general scheme outlined in the Introduction, let us perform the
Iwasawa decomposition (\ref{iwasawa},\ref{iwagroup}) of the group element
\( g_{j} \) corresponding to the discrete time \( j \) in the product of a
compact, Abelian and nilpotent part:
\begin{equation}
\label{iwa2}
g_{j}=k_{j}a_{j}n_{j}
\end{equation}
 where \( g_{j} \) is defined recursively as in (\ref{rec}) and \( k_{j} \),
\( a_{j} \), \( n_{j} \) are functions of \( g_{j} \) obtained by inverting
the diffeomorphism (\ref{diffeo}). Moreover, let us define \( K_{j} \), \( A_{j} \),
\( N_{j} \) through the relations 
\begin{equation}
\label{iwa3}
k_{j}=\exp \left( \epsilon K_{j}\right) k_{j-1},\: a_{j}=\exp \left( \epsilon A_{j}\right) a_{j-1},\: n_{j}=\exp \left( \epsilon N_{j}\right) n_{j-1}
\end{equation}
The variables \( K,A,N \) are an intermediate set of new variables, adapted
to the Iwasawa decomposition (\ref{iwa2}). In order to obtain the explicit
form of the variable transformation \( X\rightarrow \{K,A,N\} \), one substitutes
(\ref{iwa2}) in (\ref{rec}) and uses (\ref{iwa3}): 
\begin{eqnarray*}
\exp \left( \epsilon X_{j}\right)  & = & g_{j}g_{j-1}^{-1}\\
 & = & \exp \left( \epsilon K_{j}\right) \cdot \mathrm{Ad}\left( k_{j-1}\right) \exp \left( \epsilon A_{j}\right) \cdot \mathrm{Ad}\left( k_{j-1}a_{j-1}\right) \exp \left( \epsilon N_{j}\right) 
\end{eqnarray*}
 where \( \mathrm{Ad}\left( g\right) X=gXg^{-1} \). Using (\ref{midpoint})
one obtains: 
\begin{eqnarray}
X_{j} & = & \frac{1}{2\epsilon }\left[ \exp \left( \epsilon X_{j}\right) -\exp \left( -\epsilon X_{j}\right) \right] +O\left( \epsilon ^{2}X_{j}^{3}\right) \nonumber \\
 & = & K_{j}+\mathrm{Ad}\left( k_{j-1}\right) A_{j}+\mathrm{Ad}\left( k_{j-1}a_{j-1}\right) N_{j}\label{decX} \\
 &  & +\frac{\epsilon }{2}\left[ K_{j},\mathrm{Ad}\left( k_{j-1}\right) A_{j}\right] +\frac{\epsilon }{2}\left[ K_{j},\mathrm{Ad}\left( k_{j-1}a_{j-1}\right) N_{j}\right] \nonumber \\
 &  & +\frac{\epsilon }{2}\left[ \mathrm{Ad}\left( k_{j-1}\right) A_{j},\mathrm{Ad}\left( k_{j-1}a_{j-1}\right) N_{j}\right] +O\left( \epsilon ^{2}X_{j}^{3}\right) \nonumber 
\end{eqnarray}
 where \( O\left( \epsilon ^{2}X_{j}^{3}\right)  \) denotes a series of terms
of at least second order in \( \epsilon  \) and at least third order in \( X_{j} \),
\( K_{j} \), \( A_{j} \), \( N_{j} \). Using 
\[
\mathrm{Ad}\left( \exp \left( \frac{\epsilon }{2}X\right) \right) Y=Y+\frac{\epsilon }{2}\left[ X,Y\right] +O\left( \epsilon ^{2}X^{2}Y\right) \]
 this can be rewritten as 
\begin{equation}
\label{IWA}
X_{j}=K_{j}+\mathrm{Ad}\left( \tilde{k}_{j-1}\right) A_{j}+\mathrm{Ad}\left( \tilde{k}_{j-1}\tilde{a}_{j-1}\right) N_{j}+O\left( \epsilon ^{2}X_{j}^{3}\right) 
\end{equation}
 with \( \tilde{k}_{j-1}=\exp \left( \frac{\epsilon }{2}K_{j}\right) k_{j-1} \),
\( \tilde{a}_{j-1}=\exp \left( \frac{\epsilon }{2}A_{j}\right) a_{j-1} \).

Formula (\ref{IWA}) extends the Iwasawa decomposition on the Lie algebra \( \hat{g} \)
to a natural decomposition 
\[
\left( K_{j},A_{j},N_{j}\right) \rightarrow X_{j}\]
 of the random process \( X_{j} \), which corresponds to the decomposition
(\ref{iwa2}) at the group level. Observe that 
\[
\tilde{k}_{j-1}=\tilde{k}_{j-1}\left( K_{j},K_{j-1},\ldots ,K_{1}\right) ,\qquad \tilde{a}_{j-1}=\tilde{a}_{j-1}\left( A_{j},A_{j-1},\ldots ,A_{1}\right) \]
 and that the appearance of \( \tilde{k}_{j-1} \) and \( \tilde{a}_{j-1} \)
means that at the continuum level the midpoint regularization (\ref{midpoint})
applies also to the group terms arising from the Iwasawa decomposition.

\section{Decomposition of the Gibbs weight\label{quattro}}

Relation (\ref{IWA}) can now be used to express the Gibbs weight defined through
(\ref{metric}) in terms of the new variables. Using the orthogonality with
respect to \( B \) of \( \hat{a} \) to both \( \hat{k} \) and \( \hat{n} \),
of \( \hat{n} \) with itself, and properties (\ref{unomenouno},\ref{thetaad},\ref{thetaAd}),
a simple computation shows that

\begin{eqnarray}
F\left( X_{j},X_{j}\right)  & = & \phantom {+}\left( \mu +\nu \right) B\left( K_{j}-K^{0}_{j},K_{j}-K^{0}_{j}\right) \nonumber \\
 &  & +\left( \mu -\nu \right) B\left( A_{j},A_{j}\right) \label{lagr} \\
 &  & -\frac{\mu -\nu }{2}B\left( \mathrm{Ad}\left( \tilde{a}_{j-1}^{2}\right) N_{j},\theta N_{j}\right) \nonumber 
\end{eqnarray}
 with 
\[
K^{0}_{j}=-\frac{1}{2}\left[ \mathrm{Ad}\left( \tilde{k}_{j-1}\tilde{a}_{j-1}\right) N_{j}+\theta \, \mathrm{Ad}\left( \tilde{k}_{j-1}\tilde{a}_{j-1}\right) N_{j}\right] \]
 Recalling that \( B\left( X,X\right) >0 \) on \( \hat{p} \), \( B\left( X,X\right) <0 \)
on \( \hat{k} \) and \( B\left( X,\theta X\right) <0 \) on \( \hat{g} \),
and using (\ref{thetaad},\ref{thetaAd}), it is clear that \( F\left( X,X\right) >0 \)
if 
\begin{equation}
\label{positivo}
\mu >\nu \quad \textrm{and}\quad \mu <-\nu 
\end{equation}

The last two rows of (\ref{lagr}) can be written more explicitly by expanding
over the basis \( Y_{\alpha } \) defined in (\ref{base}) and a basis \( H_{l} \)
of \( \hat{a} \): 
\begin{equation}
\label{espansione}
N_{j}=\sum _{\alpha \in P}N_{j}^{\alpha }Y_{\alpha },\qquad A_{j}=\sum _{l}A_{j}^{l}H_{l}
\end{equation}
 Then using 
\begin{eqnarray}
\mathrm{ad}\left( A_{j}\right) Y_{\alpha } & = & \alpha \left( A_{j}\right) Y_{\alpha }\label{formuno} \\
\mathrm{Ad}\left( \tilde{a}_{j-1}\right) Y_{\alpha } & = & \exp \left[ \alpha \left( \exp ^{-1}\tilde{a}_{j-1}\right) \right] Y_{\alpha }\label{formdue} 
\end{eqnarray}
 with 
\begin{equation}
\label{formtre}
\exp ^{-1}\tilde{a}_{j-1}=\sum _{l=1}^{j-1}\epsilon A_{l}+\frac{\epsilon }{2}A_{j}
\end{equation}
 one gets 
\begin{eqnarray}
 &  & B\left( A_{j},A_{j}\right) =\label{baj} \\
 & = & \sum _{k,l}A_{j}^{k}A_{j}^{l}\, B\left( H_{k},H_{l}\right) \nonumber \\
 &  & B\left( \mathrm{Ad}\left( \tilde{a}_{j-1}^{2}\right) N_{j},\theta N_{j}\right) =\label{bad} \\
 & = & \sum _{\alpha ,\beta \in P}N_{j}^{\alpha }N_{j}^{\beta }\, \exp \left[ 2\alpha \left( \exp ^{-1}\tilde{a}_{j-1}\right) \right] \, B\left( Y_{\alpha },\theta Y_{\beta }\right) \nonumber 
\end{eqnarray}
 Here the matrix \( B\left( H_{k},H_{l}\right)  \) is directly related to the
Cartan matrix which characterizes the Lie algebra \( \hat{g} \). A list of
Cartan matrices for all semisimple Lie algebras can be found in Ref. \cite{helgason78}.
The matrix \( B\left( Y_{\alpha },\theta Y_{\beta }\right)  \) is non-degenerate,
it is diagonal if the multiplicities (\ref{multiplicities}) are all 1 and is
related to the way the Cartan involution \( \theta  \) permutes the roots otherwise
\cite{helgason78}.

It is often necessary to compute averages of functionals \( F\left( \left\{ a_{j}\right\} \right)  \)
which do not depend on \( k_{j} \), \( n_{j} \). In this case, the form of
(\ref{lagr}) shows that integration over \( \prod _{j=1}^{N}\mathrm{d}K_{j} \)
produces only a constant factor. Integration over \( \prod _{j=1}^{N}\mathrm{d}N_{j} \)
produces instead a factor proportional to 
\begin{eqnarray}
D_{N}^{-1/2} & \equiv  & \prod _{j=1}^{N}\prod _{\alpha \in P}\exp \left[ -\alpha \left( \exp ^{-1}\tilde{a}_{j-1}\right) \right] \nonumber \\
 & = & \prod _{j=1}^{N}\exp \left[ -2\rho \left( \exp ^{-1}\tilde{a}_{j-1}\right) \right] \label{denne} \\
 & = & \prod _{j=1}^{N}\exp \left[ -2\sum _{l=1}^{j-1}\epsilon \rho \left( A_{l}\right) -\epsilon \rho \left( A_{j}\right) \right] \nonumber 
\end{eqnarray}
 where notation (\ref{semisomma}) was used.

Expression (\ref{lagr}) takes a simpler form if one introduces the ``rotated''
variables
\begin{equation}
\label{rotated}
M_{j}=\mathrm{Ad}\left( \tilde{a}_{j-1}\right) N_{j}
\end{equation}
The variable change \( N\rightarrow M \) introduces however a Jacobian factor
\begin{eqnarray}
\frac{\partial \left( \left\{ M_{j}\right\} \right) }{\partial \left( \left\{ N_{j}\right\} \right) } & = & \prod _{j=1}^{N}\det \left( \left. \mathrm{Ad}\left( \tilde{a}_{j-1}\right) \right| _{\hat{n}}\right) \nonumber \\
 & = & \prod _{j=1}^{N}\exp \left[ \mathrm{Tr}\, \left. \mathrm{ad}\left( \exp ^{-1}\tilde{a}_{j-1}\right) \right| _{\hat{n}}\right] \nonumber \\
 & = & \prod _{j=1}^{N}\exp \left[ \sum _{\alpha \in P}\alpha \left( \exp ^{-1}\tilde{a}_{j-1}\right) \right] \nonumber \\
 & = & \prod _{j=1}^{N}\exp \left[ 2\sum _{l=1}^{j-1}\epsilon \rho \left( A_{l}\right) +\epsilon \rho \left( A_{j}\right) \right] \, =\, D_{N}^{1/2}\label{jacparz} 
\end{eqnarray}
 Integration over the variables \( M_{j} \) would than produce a constant (independent
from \( A_{j} \)) factor. Moreover, factor (\ref{jacparz}) will be seen to
cancel exactly a corresponding factor emerging in the global Jacobian (\ref{jacobiano}).

\section{Decomposition of the volume form\label{cinque}}

Along with the variables \( X_{j} \) and \( K_{j},A_{j},N_{j} \), which belong
to the algebra \( \hat{g} \) of infinitesimal transformations, one can consider
the ``integrated'' variables \( g_{j} \) and \( k_{j},a_{j},n_{j} \) defined
in (\ref{rec}) and (\ref{iwa2}). All together, one has the following commutative
diagram of variable transformations: 
\begin{eqnarray}
\left\{ g_{j}\right\}  & \longrightarrow  & \left\{ k_{j},a_{j},n_{j}\right\} \nonumber \\
\downarrow \, \,  &  & \, \, \qquad \downarrow \label{diagr} \\
\left\{ X_{j}\right\}  & \longrightarrow  & \left\{ K_{j},A_{j},N_{j}\right\} \nonumber 
\end{eqnarray}
where the variables of the upper row belong to the group \( G \) and those
of the lower row to the corresponding algebra \( \hat{g} \). The relation between
the two set of variables can be found by varying (\ref{decX}):

\begin{eqnarray}
\delta X_{j} & = & \delta K_{j}+\mathrm{Ad}\left( k_{j-1}\right) \delta A_{j}+\mathrm{Ad}\left( k_{j-1}a_{j-1}\right) \delta N_{j}\nonumber \\
 &  & +\frac{\epsilon }{2}\Big \{\left[ \delta K_{j},\mathrm{Ad}\left( k_{j-1}\right) A_{j}\right] +\left[ K_{j},\mathrm{Ad}\left( k_{j-1}\right) \delta A_{j}\right] \nonumber \\
 &  & \qquad +\left[ \delta K_{j},\mathrm{Ad}\left( k_{j-1}a_{j-1}\right) N_{j}\right] +\left[ K_{j},\mathrm{Ad}\left( k_{j-1}a_{j-1}\right) \delta N_{j}\right] \label{decdX} \\
 &  & \qquad +\left[ \mathrm{Ad}\left( k_{j-1}\right) \delta A_{j},\mathrm{Ad}\left( k_{j-1}a_{j-1}\right) N_{j}\right] \nonumber \\
 &  & \qquad +\left[ \mathrm{Ad}\left( k_{j-1}\right) A_{j},\mathrm{Ad}\left( k_{j-1}a_{j-1}\right) \delta N_{j}\right] \Big \}\nonumber \\
 &  & \qquad +O\left( \epsilon ^{2}X_{j}^{2}\delta X_{j}\right) +f\left( \delta X_{j-1},\delta X_{j-2},\ldots ,\delta X_{1}\right) \nonumber 
\end{eqnarray}
 The variation \( \delta X_{j}\in \mathrm{T}\hat{g} \) can be identified with
the 1-form \( \delta X_{j}(\, \cdot \, )=F\left( \delta X_{j},\, \cdot \, \right)  \).
In (\ref{decdX}), \( O\left( \epsilon ^{2}X_{j}^{2}\delta X_{j}\right)  \)
denotes a 1-form in \( \delta K_{j} \), \( \delta A_{j} \), \( \delta N_{j} \)
whose coefficients are at least of second order in \( \epsilon  \) and in \( K_{j} \),
\( A_{j} \), \( N_{j} \), while \( f\left( \delta X_{j-1},\delta X_{j-2},\ldots ,\delta X_{1}\right)  \)
denotes a 1-form in \( \delta K_{l} \), \( \delta A_{l} \), \( \delta N_{l} \)
for \( l<j \).

On the other hand, the measure \( \mathrm{d}\mu \left( X\right)  \) can be
re-expressed in terms of the Haar measure on \( G \), by computing the variation
\begin{eqnarray*}
\delta g_{j}g_{j}^{-1} & = & \left[ \epsilon \delta X_{j}+\frac{\epsilon ^{2}}{2}\left( \delta X_{j}\, X_{j}+X_{j}\, \delta X_{j}\right) +O\left( \epsilon ^{3}X_{j}^{2}\delta X_{j}\right) \right] \cdot g_{j-1}g_{j}^{-1}\\
 &  & +f\left( \delta X_{j-1},\ldots ,\delta X_{1}\right) \\
 & = & \left[ \epsilon \delta X_{j}+\frac{\epsilon ^{2}}{2}\left( \delta X_{j}\, X_{j}+X_{j}\, \delta X_{j}\right) \right] \left( 1-\epsilon X_{j}\right) \\
 &  & +O\left( \epsilon ^{3}X_{j}^{2}\delta X_{j}\right) +f\left( \delta X_{j-1},\ldots ,\delta X_{1}\right) \\
 & = & \epsilon \delta X_{j}+\frac{\epsilon ^{2}}{2}\left[ X_{j},\delta X_{j}\right] +O\left( \epsilon ^{3}X_{j}^{2}\delta X_{j}\right) +f\left( \delta X_{j-1},\ldots ,\delta X_{1}\right) 
\end{eqnarray*}
 so that 
\begin{eqnarray}
\frac{1}{\epsilon }\delta g_{j}g_{j}^{-1} & = & \mathrm{Ad}\left( \exp \left( \frac{\epsilon }{2}X_{j}\right) \right) \, \delta X_{j}\nonumber \\
 &  & +O\left( \epsilon ^{2}X_{j}^{2}\delta X_{j}\right) +f\left( \delta X_{j-1},\ldots ,\delta X_{1}\right) \nonumber 
\end{eqnarray}
 In the continuum limit, terms \( O\left( \epsilon ^{2}X_{j}^{2}\delta X_{j}\right)  \)
contribute to the measure with a constant factor which ensures the correct normalization\footnote{%
One can treat terms \( \sim \epsilon ^{2}X_{j}^{2}\mathrm{d}X_{j} \) as a perturbation
in integrals over the measure (\ref{discrmeas}) and use Wick's theorem to compute
their contribution. They give a contribution \( \left\langle \epsilon ^{2}X_{j}^{2}\right\rangle \sim \epsilon  \),
but their product with any power of \( \epsilon X_{j} \) higher than \( 0 \)
is negligible in the continuum limit. The contributions \( \sim \epsilon  \)
sum up to give a constant factor.\label{nota1} 
}. One then has 
\[
\bigwedge _{j=1}^{N}\bigwedge _{l=1}^{\dim \hat{g}}\delta _{l}X_{j}\propto \bigwedge _{j=1}^{N}\bigwedge _{l=1}^{\dim G}\delta _{l}g_{j}\, g_{j}^{-1}\]
 where the \( \delta _{l}X \) span \( \mathrm{T}_{0}\hat{\mathrm{g}} \), \( \delta _{l}g\, g^{-1} \)
span \( \mathrm{T}_{e}G\simeq \mathrm{T}_{0}\hat{g} \), and one uses the facts
that

\begin{enumerate}
\item the \( f\left( \delta X_{j-1},\ldots ,\delta X_{1}\right)  \) terms cancel
in the exterior product with \( \delta X_{j} \) terms; 
\item semisimple Lie groups are unimodular, \emph{i.e.}, \( \left| \det \mathrm{Ad}\left( g\right) \right| =1 \)
for every \( g\in G \). 
\end{enumerate}
So: 
\[
\prod _{j=1}^{N}\mathrm{d}X_{j}\propto \prod _{J=1}^{N}\mathrm{d}g_{j}\]
 where \( \mathrm{d}g \) denotes the Haar measure on \( G \). In other words,
the variable transformation \( \left\{ X_{j}\right\} \rightarrow \left\{ g_{j}\right\}  \)
has trivial Jacobian and \( \mathrm{d}\mu \left( X\right) =\mathrm{d}\mu \left( g\right)  \),
where \noun{
\[
\mathrm{d}\mu \left( g\right) =K'\exp \left[ -\frac{1}{2}\int _{0}^{T}F\left( \dot{g}g^{-1},\dot{g}g^{-1}\right) \mathrm{d}t\right] \prod _{t=0}^{T}\mathrm{d}g\left( t\right) \]
}

To the two horizontal arrows of diagram (\ref{diagr}) there correspond the
Jacobians defined by 
\begin{eqnarray}
\prod _{j=1}^{N}\mathrm{d}g_{j} & = & J_{N}^{G}\left( \left\{ k_{j},a_{j},n_{j}\right\} \right) \prod _{j=1}^{N}\mathrm{d}k_{j}\mathrm{d}a_{j}\mathrm{d}n_{j}\label{haar1} \\
\prod _{j=1}^{N}\mathrm{d}X_{j} & = & J_{N}^{\hat{g}}\left( \left\{ K_{j},A_{j},N_{j}\right\} \right) \prod _{j=1}^{N}\mathrm{d}K_{j}\mathrm{d}A_{j}\mathrm{d}N_{j}\label{haar2} 
\end{eqnarray}
 The variable transformations corresponding to the two vertical arrows have
trivial Jacobians, so 
\begin{equation}
\label{haar3}
J_{N}^{\hat{g}}\left( \left\{ K_{j},A_{j},N_{j}\right\} \right) \propto J_{N}^{G}\left( \left\{ k_{j},a_{j},n_{j}\right\} \right) 
\end{equation}
 The product of Haar measures on both sides of (\ref{haar1}) is invariant under
the global transformations 
\[
g_{j}\rightarrow k_{\mathrm{R}}g_{j},\qquad g_{j}\rightarrow g_{j}n_{\mathrm{L}},\qquad \textrm{with }k_{\mathrm{R}}\in K,\: n_{\mathrm{L}}\in N\]
 which correspond to 
\[
k_{j}\rightarrow k_{\mathrm{R}}k_{j},\qquad n_{j}\rightarrow n_{j}n_{\mathrm{L}}\]
 By induction over \( N \), this shows that \( J_{N}^{G} \) does not actually
depend on \( \left\{ k_{j},n_{j}\right\}  \). From (\ref{haar3}) then follows
that \( J_{N}^{\hat{g}} \) does not depend on \( \left\{ K_{j},N_{j}\right\}  \).

The actual value of the Jacobian can now be computed either at the group or
at the algebra level. At the group level, using (\ref{iwwames}), one gets 
\begin{eqnarray}
J_{N}^{G} & = & \prod _{j=1}^{N}\exp \left[ 2\rho \left( \exp ^{-1}a_{j}\right) \right] \nonumber \\
 & = & \prod _{j=1}^{N}\exp \left[ 2\sum _{l=1}^{j}\epsilon \rho \left( A_{l}\right) \right] \nonumber \\
 & = & \prod _{j=1}^{N}\exp \left[ 2\sum _{l=1}^{j-1}\epsilon \rho \left( A_{l}\right) +\epsilon \rho \left( A_{j}\right) \right] \cdot \exp \left[ \sum _{j=1}^{N}\epsilon \rho \left( A_{j}\right) \right] \nonumber \\
 & = & D_{N}^{1/2}\cdot J_{N}^{G,\mathrm{loc}}\label{jacobiano} 
\end{eqnarray}
 The decomposition of \( J^{G}_{N} \) into an ultralocal part \( D_{N}^{1/2} \)
and a local part \( J_{N}^{G,\mathrm{loc}} \) is such that the ultralocal part
cancels exactly the factor (\ref{denne}) which appears when integrating functionals
of the form \( F\left( \left\{ a_{j}\right\} \right)  \) over \( \prod _{j=1}^{N}\mathrm{d}K_{j}\cdot \prod _{j=1}^{N}\mathrm{d}N_{j} \).
The ultralocal factor \( D_{N}^{1/2} \) is also exactly canceled when passing
to the ``rotated'' variables \( M_{j} \) defined in Eq. (\ref{rotated}). 

The result (\ref{jacobiano}) can be checked by repeating the computation at
the algebra level. Recalling that \( J_{N}^{\hat{g}} \) cannot depend on \( K_{j} \),
\( N_{j} \), we can set \( K_{j}=0 \), \( N_{j}=0 \) in (\ref{decdX}), getting

\begin{eqnarray}
\delta X_{j} & = & \delta K_{j}+\delta A_{j}+\mathrm{Ad}\left( a_{\mathrm{j}-1}\right) \delta N_{j}\nonumber \\
 &  & +\frac{\epsilon }{2}\left[ \delta K_{j},A_{j}\right] +\frac{\epsilon }{2}\left[ A_{j},\mathrm{Ad}\left( a_{j-1}\right) \delta N_{j}\right] \nonumber \\
 &  & +O\left( \epsilon ^{2}X_{j}^{2}\delta X_{j}\right) +f\left( \delta X_{j-1},\ldots ,\delta X_{1}\right) \label{dxj} \\
 & = & \mathrm{Ad}\left( \exp \left( -\frac{\epsilon }{2}A_{j}\right) \right) \, \delta K_{j}+\delta A_{j}+\mathrm{Ad}\left( \exp \left( \frac{\epsilon }{2}A_{j}\right) \, a_{j-1}\right) \delta N_{j}\nonumber \\
 &  & +O\left( \epsilon ^{2}X_{j}^{2}\delta X_{j}\right) +f\left( \delta X_{j-1},\ldots ,\delta X_{1}\right) \nonumber 
\end{eqnarray}
 Using (\ref{kesp}) and (\ref{base}) to expand \( N_{j}=\sum _{\alpha \in P}\delta N_{j}^{\alpha }Y_{\alpha } \),
one gets 
\begin{eqnarray}
 &  & \mathrm{Ad}\left( \exp \left( \frac{\epsilon }{2}A_{j}\right) \, a_{j-1}\right) \delta N_{j}\nonumber \\
 & = & \sum _{\alpha \in P}\exp \left[ \alpha \left( \frac{\epsilon }{2}A_{j}+\sum _{l=1}^{j-1}\epsilon A_{j}\right) \right] \delta N_{j}^{\alpha }Y_{\alpha }\label{formenne} 
\end{eqnarray}
 Using (\ref{kesp}) and (\ref{base}) to expand \( \delta K_{j}=\delta K_{j}^{0}+\sum _{\alpha \in P}\delta K_{j}^{\alpha }\left( Y_{\alpha }+\theta Y_{\alpha }\right)  \),
where \( \delta K_{j}^{0}\in \hat{m} \), and recalling (\ref{cartancambia}),
one gets 
\begin{eqnarray*}
\mathrm{ad}\left( A_{j}\right) \delta K_{j} & = & \sum _{\alpha \in P}\alpha \left( A_{j}\right) \delta K^{\alpha }_{J}\left( Y_{\alpha }-\theta Y_{\alpha }\right) \\
 & = & \sum _{\alpha \in P}\alpha \left( A_{j}\right) \delta K^{\alpha }_{J}\left[ 2Y_{\alpha }-\left( Y_{\alpha }+\theta Y_{\alpha }\right) \right] \\
 & = & N^{0}_{j}-\sum _{\alpha \in P}\alpha \left( A_{j}\right) \delta K^{\alpha }_{J}\left( Y_{\alpha }+\theta Y_{\alpha }\right) 
\end{eqnarray*}
 with \( N^{0}_{j}\in \hat{n} \), and 
\begin{eqnarray}
\left( 1-\frac{\epsilon }{2}\mathrm{ad}\left( A_{j}\right) \right) \delta K_{j} & = & \sum _{\alpha \in P}\exp \left[ \alpha \left( \frac{\epsilon }{2}A_{j}\right) \right] \delta K^{\alpha }_{j}\left( Y_{\alpha }+\theta Y_{\alpha }\right) \nonumber \\
 &  & -\frac{\epsilon }{2}N^{0}_{j}+O\left( \epsilon ^{2}A_{j}^{2}\delta K_{j}\right) \label{formkappa} 
\end{eqnarray}
 Now using (\ref{dxj},\ref{formenne},\ref{formkappa}) and the decomposition
(\ref{iwasawa}), observing that the terms \( f\left( \delta X_{j-1},\ldots ,\delta X_{1}\right)  \)
cancel in the exterior product with \( \delta X_{j} \) (the transformation
\( \left\{ X_{j}\right\} \rightarrow \left\{ K_{j},A_{j},N_{j}\right\}  \)
is triangular in the index \( j \)), and that \( O\left( \epsilon ^{2}X_{j}^{2}\delta X_{j}\right)  \)
terms contribute only to a normalization factor (see footnote \ref{nota1}),
one gets 
\begin{eqnarray}
J^{\hat{g}}_{N} & \propto  & \prod _{j=1}^{N}\prod _{\alpha \in P}\exp \left[ \alpha \left( \sum _{l=1}^{j-1}\epsilon A_{j}+\frac{\epsilon }{2}A_{j}\right) \right] \exp \left[ \alpha \left( \frac{\epsilon }{2}A_{j}\right) \right] \nonumber \\
 & = & \prod _{j=1}^{N}\exp \left[ 2\sum _{l=1}^{j-1}\epsilon \rho \left( A_{l}\right) +\epsilon \rho \left( A_{j}\right) \right] \cdot \exp \left[ \sum _{j=1}^{N}\epsilon \rho \left( A_{j}\right) \right] \nonumber \\
 & = & D^{1/2}\cdot J^{\hat{g},\mathrm{loc}}_{\mathrm{N}}\label{glocjac} 
\end{eqnarray}
 as expected.

Taking into account (\ref{jacobiano}), (\ref{glocjac}) and (\ref{jacparz})
it is clear that the Jacobian of the transformation 
\[
\left\{ X_{j}\right\} \rightarrow \left\{ K_{j},A_{j},M_{j}\right\} \]
 is proportional to the simple local factor 
\begin{equation}
\label{locfac}
J^{\hat{g},\mathrm{loc}}_{\mathrm{N}}=\exp \left[ \sum _{j=1}^{N}\epsilon \rho \left( A_{j}\right) \right] 
\end{equation}

\section{Statistics of Lyapunov exponents\label{sei}}

Classical groups are groups of transformations of \( \mathbf{R}^{n} \) or \( \mathbf{C}^{n} \).
Let \( (\, ,\, ) \) be a scalar product on \( \mathbf{R}^{n} \) or a hermitian
product on \( \mathbf{C}^{n} \). Let \( \mathbf{v} \) be a vector of \( \mathbf{R}^{n} \)
or \( \mathbf{C}^{n} \), \( g\mathbf{v} \) the action of the group element
\( g \) on \( \mathbf{v} \). Let \( g\left( t\right)  \) be a random walk
on a (real or complex) classical group \( G \). Then 
\[
\left\Vert g\left( t\right) \mathbf{v}\right\Vert ^{2}=\left( g\left( t\right) \mathbf{v}\, ,\, g\left( t\right) \mathbf{v}\right) =\left( g^{\dagger }\left( t\right) g\left( t\right) \mathbf{v}\, ,\, \mathbf{v}\right) \]
 so that the rate of growth of \( \left\Vert g\left( t\right) \mathbf{v}\right\Vert  \)
is related to the highest eigenvalue of \( g^{\dagger }\left( t\right) g\left( t\right)  \).
More generally, the rate of growth of the parallelepiped generated by \( l \)
vectors \( \mathbf{v}_{1},\ldots ,\mathbf{v}_{l} \), is related to the sum
of the first \( l \) eigenvalues \( \exp \left( 2\lambda _{1}\left( t\right) t\right) ,\ldots ,\exp \left( 2\lambda _{l}\left( t\right) t\right)  \)
of the hermitian matrix \( g^{\dagger }\left( t\right) g\left( t\right)  \)
\cite{cpv93}: 
\[
\lambda _{1}\left( t\right) +\cdots +\lambda _{l}\left( t\right) =\frac{1}{t}\log \, \frac{\mathrm{Vol}\left( g\left( t\right) \mathbf{v}_{1}\wedge \cdots \wedge g\left( t\right) \mathbf{v}_{l}\right) }{\mathrm{Vol}\left( \mathbf{v}_{1}\wedge \cdots \wedge \mathbf{v}_{l}\right) }\]
 The eigenvalues \( \lambda _{1}\left( t\right) ,\ldots ,\lambda _{l}\left( t\right)  \)
are the (time dependent) Lyapunov exponents of the process \( g\left( t\right)  \).

For the classical algebras the Cartan involution \( \theta  \) is just \( \theta Z=-Z^{\dagger } \),
which exponentiates to the group automorphism \( \Theta  \) defined by \( \Theta g=\left( g^{\dagger }\right) ^{-1} \).
Let \( T \) be a finite dimensional representation of \( G \) such that (\ref{ultimocartan})
holds. In general, the (time dependent) Lyapunov exponents of the process \( g\left( t\right)  \)
can be seen as the eigenvalues of 
\[
\left( 2t\right) ^{-1}\, \log \left[ T^{\dagger }\left( g\right) T\left( g\right) \right] =-\left( 2t\right) ^{-1}\, \log T\left( g^{-1}\Theta g\right) =\left( 2t\right) ^{-1}\, \log T\left( \Theta n^{-1}\cdot a^{2}\, n\right) \]
 or equivalently as the eigenvalues of 
\[
\left( 2t\right) ^{-1}\log \left[ T\left( a^{2}\right) T\left( n\Theta n^{-1}\right) \right] \]
 After integrating out the \( \hat{k} \) component, incorporating the contribution
of \( J_{N}^{\hat{g},\mathrm{loc}} \), and using (\ref{baj},\ref{bad},\ref{bhh}),
\( F\left( X_{j},X_{j}\right)  \) reduces to ,
\begin{eqnarray}
 &  & F^{\hat{a}+\hat{n}}\, \left( X_{j},X_{j}\right) =\nonumber \\
 & = & \left( \mu -\nu \right) \sum _{k,l}B\left( H_{k},H_{l}\right) \, A_{j}^{k}A_{j}^{l}-\sum _{l}\sum _{\alpha \in P}B\left( H_{l},H_{\alpha }\right) \, A^{l}_{j}\nonumber \\
 &  & -\frac{\mu -\nu }{2}\sum _{\alpha ,\beta \in P}\exp \left[ \alpha \left( \sum _{l=1}^{j-1}2\epsilon A_{j}+\epsilon A_{j}\right) \right] \, B\left( Y_{\alpha },\theta Y_{\beta }\right) \, N_{j}^{\alpha }N_{j}^{\beta }\label{gred} \\
 & = & F^{\hat{a}}\, \left( X_{j},X_{j}\right) -\frac{\mu -\nu }{2}\sum _{\alpha ,\beta \in P}B\left( Y_{\alpha },\theta Y_{\beta }\right) \, M_{j}^{\alpha }M_{j}^{\beta }\label{gred2} 
\end{eqnarray}
 After integrating out the \( M \)-variables, and taking into account the observations
following (\ref{denne}) and (\ref{jacobiano}), \( F^{\hat{a}+\hat{n}}\, \left( X_{j},X_{j}\right)  \)
reduces to \( F^{\hat{a}}\, \left( X_{j},X_{j}\right)  \). This shows that
in the \( \left\{ K_{j},A_{j},M_{j}\right\}  \) representation the \( A_{j} \)
are Gaussian random variables with positive mean 
\begin{equation}
\label{media}
\left\langle A_{j}^{k}\right\rangle =\frac{1}{2\left( \mu -\nu \right) }\sum _{l}\Gamma ^{kl}\, \sum _{\alpha \in P}B\left( H_{l},H_{\alpha }\right) 
\end{equation}
 where 
\begin{equation}
\label{defcov}
\left( \Gamma ^{-1}\right) _{kl}=B\left( H_{k},H_{l}\right) 
\end{equation}
 and covariance 
\begin{equation}
\label{covarianza}
\left\langle A_{j}^{k}A_{j}^{l}\right\rangle _{\mathrm{c}}=\frac{1}{\mu -\nu }\Gamma ^{kl}
\end{equation}
 On the other hand, (\ref{media},\ref{covarianza}) together with (\ref{gred})
show that the \( n \) variables diffuse, while the \( a \) variables perform
a ballistic motion, diffusing around straight trajectories. This means that
for \( t\gg 1 \) and with logarithmic precision the time dependent Lyapunov
exponents coincide with the eigenvalues of 
\begin{eqnarray}
 &  & \frac{1}{2t}\, \log T\left( a^{2}\left( t\right) \right) \nonumber \\
 & = & \frac{1}{t}\, \tau \left( \exp ^{-1}\left( a\left( t\right) \right) \right) \nonumber \\
 & = & \frac{1}{t}\, \tau \left( \int _{0}^{t}A\left( t'\right) \mathrm{d}t'\right) \nonumber \\
 & = & \frac{1}{t}\int _{0}^{t}\mathrm{d}t'\, \sum _{k}A^{k}\left( t'\right) \, \tau _{k}\label{blocco} 
\end{eqnarray}
 with \( T\left( \exp \left( X\right) \right) =\exp \left( \tau \left( X\right) \right)  \)
and \( \tau _{k}=\tau \left( H_{k}\right)  \). The operators \( \tau _{k} \)
all commute, so it is possible to write (\ref{blocco}) in block-matrix form.
This means that the Lyapunov exponents are just the \( \frac{1}{t}\int _{0}^{t}\mathrm{d}t'\, A^{k} \)
multiplied by the eigenvalues of the operators \( \tau _{k} \).

Let us describe more precisely the statistics of the \( n \) variables. Expression
(\ref{gred2}) shows that the \( M_{j} \) are Gaussian random variables with
zero mean and \( O\left( 1\right)  \) covariance. Expanding the \( M_{j} \)
and \( N_{j} \) on the basis \( Y_{\alpha } \) one finds the explicit expression
\begin{equation}
\label{esprexp}
N^{\alpha }_{j}=\exp \left[ -\alpha \left( \sum _{l=1}^{j-1}\epsilon A_{l}+\frac{\epsilon }{2}A_{j}\right) \right] M_{j}^{\alpha },\qquad \alpha \in P
\end{equation}
 From the positiveness of \( \alpha  \) and of \( \left\langle A_{j}^{k}\right\rangle  \),
\( k=1,\ldots ,\dim \hat{g} \), and passing to the continuum limit, it follows
that typical values of the \( N^{\alpha }\left( t\right)  \) are exponentially
small in \( t \). Correspondingly, the \( n\left( t\right)  \) variables become
frozen after a time \( t=O\left( 1\right)  \). The freezing of the \( n \)
variables has been observed in Ref. \cite{cfk98} and used to obtain analytic
results about the statistics of the dissipation of a passive scalar field in
Refs. \cite{cfk98,gk99}. Moreover, it provides a nice realization of Oseledec
theorem about the stability of the Lyapunov basis in the \( t\rightarrow \infty  \)
limit \cite{oseledec68}.

\section{Examples\label{sette}}

Let \( E_{ij} \) denote a square matrix with entry \( 1 \) where the \( i \)th
row and the \( j \)th column meet, all other entries being \( 0 \). Then

\[
E_{ij}E_{kl}=\delta _{jk}E_{il},\qquad \left[ E_{ij},E_{kl}\right] =\delta _{jk}E_{il}-\delta _{li}E_{kj}\]

\paragraph*{The case of \protect\( \hat{s}\hat{l}\left( n,\mathbf{R}\right) \protect \). }

In this case \( \hat{g}_{\mathbf{C}} \) is the algebra of \( n\times n \)
matrices with zero trace. Letting 
\begin{equation}
\label{def1}
H_{i}=E_{ii}-E_{nn}\quad \left( 1\leq i\leq n-1\right) ,\qquad \hat{h}_{\mathbf{C}}=\sum _{i}\mathbf{C}H_{i}
\end{equation}
 one has the direct decomposition \cite{helgason78}
\begin{equation}
\label{decomp}
\hat{g}_{\mathbf{C}}=\hat{h}_{\mathbf{C}}+\sum _{i\ne j}\mathbf{C}E_{ij}
\end{equation}
 If \( H\in \hat{h}_{\mathbf{C}} \) and \( e_{i}\left( H\right)  \) (\( 1\leq i\leq n \))
are the diagonal elements of \( H \), one has 
\begin{equation}
\label{action}
\left[ H,E_{ij}\right] =\left( e_{i}\left( H\right) -e_{j}\left( H\right) \right) E_{ij}
\end{equation}
 The Killing form is 
\begin{equation}
\label{kill1}
B\left( X,Y\right) =2n\mathrm{Tr}\left( XY\right) 
\end{equation}
 The roots are 
\[
e_{i}-e_{j}\qquad \left( 1\leq i,j\leq n\right) \]
 which are duals to the elements \( \frac{1}{2n}\left( H_{i}-H_{j}\right) \in \hat{h}_{\mathbf{R}} \).
The compact real form for \( \hat{s}\hat{l}\left( n,\mathbf{C}\right)  \) is
\( \hat{u}=\hat{s}\hat{u}\left( n\right)  \), the algebra of skew hermitian
matrices; \( \hat{h}_{\mathbf{C}} \) are the diagonal matrices with null trace,
\( \hat{a}=\hat{h}_{\mathbf{R}} \) are the real matrices of \( \hat{h}_{\mathbf{C}} \).
Taking lexicographic ordering with respect to the basis \( e_{i}-e_{n} \) \( \left( 1\leq i\leq n-1\right)  \)
of the dual of \( \hat{h}_{\mathbf{R}} \), \emph{i.e.} assuming \( e_{i}-e_{n}>e_{j}-e_{n} \)
and consequently \( e_{i}-e_{j}>0 \) for \( i<j \), one finds 
\[
\hat{n}_{\mathbf{C}}=\sum _{i<j}\mathbf{C}E_{ij}=\left\{ \textrm{strictly upper triangular matrices}\right\} \]
 \( \hat{g}=\hat{s}\hat{l}\left( n,\mathbf{R}\right)  \) is the real form corresponding
to the conjugation \( \sigma Z=\bar{Z} \). The Cartan involution is \( \theta X=-X^{\mathrm{tr}} \).
One has \( \hat{k}=\hat{u}\cap \hat{g}=\hat{s}\hat{o}\left( n\right)  \), \( \hat{p} \)
are the real symmetric matrices with null trace, \( \hat{a} \) the diagonal
matrices of \( \hat{p} \), \( \hat{n} \) the real matrices of \( \hat{n}_{\mathbf{C}} \).
The restricted roots coincide with the roots \( e_{i}-e_{j} \). Letting \( A_{j}=\sum _{k=1}^{n-1}A^{k}_{j}H_{k} \)
and using (\ref{media},\ref{defcov},\ref{covarianza},\ref{def1},\ref{kill1}),
a simple computation gives: 
\begin{equation}
\label{mediasl}
\left\langle A^{k}_{j}\right\rangle =\frac{1}{2\left( \mu -\nu \right) }\frac{n-2k+1}{2n},\qquad k=1,\ldots ,n-1
\end{equation}
 and 
\[
\left\langle A^{k}_{j}A^{l}_{j}\right\rangle _{\mathrm{c}}=\frac{1}{2n\left( \mu -\nu \right) }\left( \delta ^{kl}-\frac{1}{n}\right) \]
 Eq. (\ref{mediasl}) reproduces Eq. (39) in Ref. \cite{gk96} letting \( \mu =1/4Dn \),
\( \nu =0 \) and Eq. (25) in Ref. \cite{gk99} letting \( \mu =1/2n^{2}\left( n+2\right) D \),
\( \nu =-\left( n+1\right) /2n^{2}\left( n+2\right) D \).

\paragraph*{The case of \protect\( \hat{s}\hat{p}\left( n,\mathbf{R}\right) \protect \). }

In this case \( \hat{g}_{\mathbf{C}} \) is the algebra of complex matrices
of the form \( \left( \begin{array}{cc}
X & Y\\
Z & -X^{\mathrm{tr}}
\end{array}\right)  \), with \( Y=Y^{\mathrm{tr}} \) and \( Z=Z^{\mathrm{tr}} \). Letting 
\begin{equation}
\label{defsp}
H_{i}=E_{ii}-E_{n+i,n+i}\quad \left( 1\leq i\leq n\right) ,\qquad \hat{h}_{\mathbf{C}}=\sum _{i=1}^{n}\mathbf{C}H_{i}
\end{equation}
 one has the direct decomposition \cite{helgason78}
\begin{eqnarray}
\hat{g}_{\mathbf{C}} & = & \hat{h}_{\mathbf{C}}+\sum _{i\leq j}\mathbf{C}\left( E_{n+i,j}+E_{n+j,i}\right) +\sum _{i\leq j}\mathbf{C}\left( E_{i,n+j}+E_{j,n+i}\right) \nonumber \\
 &  & \quad +\sum _{i\ne j}\mathbf{C}\left( E_{i,j}+E_{n+j,n+i}\right) \label{decompsp} 
\end{eqnarray}
 Let \( e_{j} \) be the linear form on \( \hat{h} \) given by \( e_{j}\left( H_{i}\right) =\delta _{ij} \);
then 
\begin{eqnarray*}
\left[ H,E_{n+i,j}+E_{n+j,i}\right]  & = & -\left( e_{i}\left( H\right) +e_{j}\left( H\right) \right) \left( E_{n+i,j}+E_{n+j,i}\right) \qquad \left( i\leq j\right) \\
\left[ H,E_{i,n+j}+E_{j,n+i}\right]  & = & \left( e_{i}\left( H\right) +e_{j}\left( H\right) \right) \left( E_{i,n+j}+E_{j,n+i}\right) \qquad \left( i\leq j\right) \\
\left[ H,E_{i,j}-E_{n+j,n+i}\right]  & = & \left( e_{i}\left( H\right) -e_{j}\left( H\right) \right) \left( E_{i,j}-E_{n+j,n+i}\right) \qquad \left( i\ne j\right) 
\end{eqnarray*}
 The Killing form is 
\begin{equation}
\label{killsp}
B\left( X,Y\right) =\left( 2n+2\right) \mathrm{Tr}\left( XY\right) 
\end{equation}
 The non zero roots are (\( \pm  \) signs read independently) 
\[
\pm 2e_{i}\quad \left( 1\leq 1\leq n\right) ,\qquad \pm e_{i}\pm e_{j}\quad \left( 1\leq i<j\leq n\right) \]
 The roots \( 2e_{j} \) are dual to the elements \( \frac{2}{4\left( n+1\right) }H_{i}\in \hat{h}_{\mathbf{R}} \).
The compact real form is

\[
\hat{u}=\hat{s}\hat{p}\left( n\right) =\left\{ \left( \begin{array}{cc}
U & V\\
-\bar{V} & \bar{U}
\end{array}\right) :\, U+U^{\dagger }=0,\, V=V^{\mathrm{tr}}\right\} \]
 Taking lexicographic ordering with respect to the basis \( e_{i} \) (\( 1\leq i\leq n \))
of the dual of \( \hat{h}_{\mathbf{R}} \) it is seen that the positive roots
are \( e_{i}+e_{j} \) (\( i\leq j \)) and \( e_{i}-e_{j} \) (\( i<j \)).
Consequently 
\begin{eqnarray*}
\hat{n}_{\mathbf{C}} & = & \sum _{i\leq j}\mathbf{C}\left( E_{i,n+j}+E_{j,n+i}\right) +\sum _{i<j}\mathbf{C}\left( E_{i,j}-E_{n+j,n+i}\right) \\
 & = & \left\{ \left( \begin{array}{cc}
X & Y\\
0 & -X^{\mathrm{tr}}
\end{array}\right) ,\, Y=Y^{\mathrm{tr}},\, X\textrm{ strictly upper triangular}\right\} 
\end{eqnarray*}
 \( \hat{g}=\hat{s}\hat{p}\left( n,\mathbf{R}\right)  \) is the real form corresponding
to the conjugation \( \sigma Z=\bar{Z} \) . One has 
\begin{eqnarray*}
\hat{k} & = & \hat{s}\hat{p}\left( n\right) \cap \hat{s}\hat{o}\left( 2n\right) \\
\hat{p} & = & \left\{ \left( \begin{array}{cc}
X & Y\\
Y & -X
\end{array}\right) :\, X=X^{\mathrm{tr}},\, X=\bar{X},\, Y=Y^{\mathrm{tr}},\, Y=\bar{Y}\right\} \\
\hat{a} & = & \left\{ \textrm{diagonal matrices of }\hat{p}\right\} \\
\hat{n} & = & \left\{ \textrm{real matrices of }\hat{n}_{\mathbf{C}}\right\} 
\end{eqnarray*}
 Restricted roots just coincide with roots. Letting \( A_{j}=\sum _{k=1}^{n}A^{k}_{j}H_{k} \)
and using (\ref{media},\ref{defcov},\ref{covarianza},\ref{defsp},\ref{killsp}),
a simple computation gives

\[
\left\langle A^{k}_{j}\right\rangle =\frac{1}{2\left( \mu -\nu \right) }\frac{n-k+1}{2\left( n+1\right) },\qquad k=1,\ldots ,n\]
 and 
\[
\left\langle A^{k}_{j}A^{l}_{j}\right\rangle _{\mathrm{c}}=\frac{1}{4\left( n+1\right) \left( \mu -\nu \right) }\delta ^{kl}\]

\subsubsection*{Acknowledgement}

Discussions with Igor Kolokolov, Roberto Camporesi and Lamberto Rondoni are
gratefully acknowledged. This work was supported by MURST ``Cofin 2001'', CNR
``Short term mobility 2001'' and INdAM-GNFM. 

\appendix

\section{Lie-Cartan theory\label{liecartan}}

I recall here some results from the theory of Lie algebras \cite{helgason78,helgason84,knapp86,knapp96}
for convenience and in order to establish the notations, which partially parallel
those of Ref. \cite{helgason78} .

Any complex Lie algebra \( \hat{g}_{\mathbf{C}} \) can be seen as a real Lie
algebra endowed with a complex structure. A real form of a complex Lie algebra
\( \hat{g}_{\mathbf{C}} \) is a real Lie algebra \( \hat{g} \) such that \( \hat{g}_{\mathbf{C}}=\hat{g}+i\hat{g} \)
(direct sum of vector spaces) where \( \hat{g} \) and \( i\hat{g} \) are identified
with subalgebras of \( \hat{g}_{\mathbf{C}} \). Thus \( \hat{g}_{\mathbf{C}} \)
is isomorphic to the complexification of \( \hat{g} \). (\emph{E.g.}, \( \hat{s}\hat{l}\left( n,\mathbf{R}\right)  \)
is a real form of \( \hat{s}\hat{l}\left( n,\mathbf{C}\right)  \)). There is
a 1-1 correspondence between the real forms of a Lie algebra and the conjugations
\( \sigma :X+iY\rightarrow X-iY \) (\( X,\, Y\in \hat{g} \)). (\emph{E.g.},
\( \hat{s}\hat{l}\left( n,\mathbf{R}\right)  \) corresponds to the conjugation
\( \sigma :Z\rightarrow \bar{Z} \) of \( \hat{s}\hat{l}\left( n,\mathbf{C}\right)  \)).
A compact Lie algebra is a real Lie algebra which is isomorphic to the Lie algebra
of a compact Lie group. Every semisimple Lie algebra \( \hat{g}_{\mathbf{C}} \)
over \( \mathbf{C} \) has a compact real form \( \hat{u} \), which is unique
(up to automorphisms; \emph{e.g.}, \( \hat{s}\hat{u}\left( n\right)  \) is
the compact real form of \( \hat{s}\hat{l}\left( n,\mathbf{C}\right)  \); it
corresponds to the conjugation \( \tau :Z\rightarrow -Z^{\dagger } \)).

A Cartan subalgebra \( \hat{h}_{\mathbf{C}} \) of a complex Lie algebra \( \hat{g}_{\mathbf{C}} \)
is a maximal Abelian subalgebra such that the endomorphism \( \mathrm{ad}\, H:X\rightarrow \left[ H,X\right]  \)
is semisimple (\emph{i.e.}, \( \hat{g}_{\mathbf{C}} \) can be written as a
direct sum of irreducible invariant subspaces of \( \mathrm{ad}\, H \) for
all \( H\in \hat{h}_{\mathbf{C}} \)). Every semisimple Lie algebra over \( \mathbf{C} \)
has a Cartan subalgebra which is unique (up to automorphisms; \emph{e.g.}, diagonal
matrices with null trace form a Cartan subalgebra of \( \hat{s}\hat{l}\left( n,\mathbf{C}\right)  \)).

On each real and complex Lie algebra one can define the Killing form \( B\left( X,Y\right) =\mathrm{Tr}\left( \mathrm{ad}\, X\, \mathrm{ad}\, \mathrm{Y}\right)  \),
where \( \mathrm{Tr} \) denotes the trace of the vector space endomorphism.
(\emph{E.g.}, \( B\left( X,Y\right) =2n\mathrm{Tr}\, XY \) for \( X,\, Y\in \hat{s}\hat{l}\left( n,\mathbf{C}\right)  \)).
Every semisimple Lie algebra over \( \mathbf{C} \) has a real form \( \hat{u} \)
such that \( B \) is strictly negative definite on \( \hat{u} \) (it then
follows that \( \hat{u} \) is compact, \emph{cf.} above; \emph{e.g. \( \mathrm{Tr}\, \mathrm{X}^{2}=-\mathrm{Tr}\, \mathrm{XX}^{\dagger }<0 \)}
for \( X\in \hat{s}\hat{u}\left( n\right)  \), \( X\ne 0 \)).

A complex Lie algebra \( \hat{g}_{\mathbf{C}} \) over \( \mathbf{C} \) can
be decomposed in the direct sum of the simultaneous eigenspaces (root subspaces)
\[
\hat{g}_{\mathbf{C}}^{\alpha }=\left\{ X\in \hat{g}_{\mathbf{C}}\, :\, \mathrm{ad}(H)\, X=\alpha \left( H\right) X\, \textrm{for all }H\in \hat{h}_{\mathbf{C}}\right\} \]
 where the linear function \( \alpha :\hat{h}_{\mathbf{C}}\rightarrow \mathbf{C} \)
is called a root (of \( \hat{g}_{\mathbf{C}} \) with respect to \( \hat{h}_{\mathbf{C}} \))
if \( \hat{g}_{\mathbf{C}}^{\alpha }\ne \left\{ 0\right\}  \). Let \( \Delta  \)
denote the set of nonzero roots. One has the root decomposition 
\begin{equation}
\label{rootdec}
\hat{g}_{\mathbf{C}}=\hat{h}_{\mathbf{C}}+\sum _{\alpha \in \Delta }\hat{g}_{\mathbf{C}}^{\alpha }\qquad (\textrm{direct sum})
\end{equation}
 with the properties: \emph{i) \( \mathrm{dim}\, \hat{g}_{\mathbf{C}}^{\alpha }=1 \)}
for each \( \alpha \in \Delta  \); \emph{ii)} \( \hat{g}_{\mathbf{C}}^{\alpha } \)
and \( \hat{g}_{\mathbf{C}}^{\beta } \) are orthogonal under \( B \) if \( \alpha  \),
\( \beta  \) are any two roots and \( \beta \ne -\alpha  \); \emph{iii) \( B \)}
is non degenerate on \( \hat{h}_{\mathbf{C}} \) and for each linear function
\( \alpha  \) on \( \hat{h}_{\mathbf{C}} \) there exists a unique \( H_{\alpha }\in \hat{h}_{\mathbf{C}} \)
such that 
\begin{equation}
\label{bhh}
\alpha \left( H\right) =B\left( H,H_{\alpha }\right) 
\end{equation}
 for all \( H\in \hat{h}_{\mathbf{C}} \); \emph{iv)} if \( \alpha \in \Delta  \),
then \( -\alpha \in \Delta  \) and \( \left[ \hat{g}_{\mathbf{C}}^{\alpha },\hat{g}_{\mathbf{C}}^{-\alpha }\right] =\mathbf{C}H_{\alpha } \)
and \( \alpha \left( H_{\alpha }\right) \ne 0 \). Let \( \hat{h}_{\mathbf{R}}=\sum _{\alpha \in \Delta }\mathbf{R}H_{\alpha } \).
Then by \emph{iii)}, \( \Delta  \) can be seen as a subset of the dual space
\( \hat{h}_{\mathbf{R}}' \). Given any basis on \( \hat{h}_{\mathbf{R}} \)
one can introduce a lexicographic order on \( \hat{h}_{\mathbf{R}}' \) and
thus on \( \Delta  \). This gives the concept of positive and negative roots.
A positive root is called simple if it cannot be written as the sum of two positive
roots. There are exactly \( \dim \, \hat{h}_{\mathbf{R}} \) simple roots \( \alpha _{j} \)
and each \( \alpha \in \Delta  \) can be written as \( \sum n_{j}\alpha _{j} \)
with \( n_{j} \) integers which are either all positive or all negative.

For each \( \alpha \in \Delta  \) one can choose \( X_{\alpha }\in \hat{g}_{\mathbf{C}}^{\alpha } \)
such that for all \( \alpha  \), \( \beta \in \Delta  \): 
\begin{eqnarray*}
\left[ X_{\alpha },X_{-\alpha }\right]  & = & H_{\alpha }\\
\left[ H,X_{\alpha }\right]  & = & \alpha \left( H\right) X_{\alpha }\textrm{ for }H\in \hat{h}_{\mathbf{C}}\\
\left[ X_{\alpha },X_{\beta }\right]  & = & 0\textrm{ if }\alpha \ne -\beta \textrm{ and }\alpha +\beta \not \in \Delta \\
\left[ X_{\alpha },X_{\beta }\right]  & = & N_{\alpha ,\beta }X_{\alpha +\beta }\textrm{ if }\alpha +\beta \in \Delta ,\textrm{ with }N_{\alpha ,\beta }=-N_{-\alpha ,-\beta }\\
B\left( X_{\alpha },X_{\beta }\right)  & = & \delta _{\alpha +\beta ,0}
\end{eqnarray*}
 On the real algebra \( \hat{u} \) generated by the elements \( iH_{\alpha } \),
\( X_{\alpha }-X_{-\alpha } \), \( i\left( X_{\alpha }+X_{-\alpha }\right)  \),
\( \alpha \in \Delta  \), the form \( B \) is strictly negative definite;
so, \( \hat{u} \) is a compact real form of \( \hat{g}_{\mathbf{C}} \) (\emph{cf.}
above). If \( \tau  \) is the conjugation of \( \hat{g}_{\mathbf{C}} \) with
respect to \( \hat{u} \), \( \tau X_{\alpha }=-X_{-\alpha } \) for \( \alpha \in \Delta  \).
The set \( \left\{ X_{\alpha }\right\}  \) is called a Weyl basis.

A Cartan decomposition of a semisimple Lie algebra \( \hat{g} \) over \( \mathbf{R} \)
is a direct sum \( \hat{g}=\hat{k}+\hat{p} \) in a subalgebra \( \hat{k} \)
and a vector subspace \( \hat{p} \) such that the mapping \( \theta :K+P\rightarrow K-P \)
(\( K\in \hat{k} \), \( P\in \hat{p} \)) is an automorphism of \( \hat{g} \)
and the form \( -B(X,\theta Y) \) is strictly positive definite. \( \theta  \)
is called a Cartan involution. Each semisimple Lie algebra over \( \mathbf{R} \)
has a Cartan decomposition. If \( \hat{g}_{\mathbf{C}} \) is the complexification
of \( \hat{g} \), \( \sigma  \) is the conjugation of \( \hat{g}_{\mathbf{C}} \)
with respect to \( \hat{g} \), one can always choose a real form \( \hat{u} \)
such that \( \sigma \hat{u}\subset \hat{u} \), so that \( \hat{k}=\hat{u}\cap \hat{g} \),
\( \hat{p}=i\hat{u}\cap \hat{g} \) and \( \hat{u}=\hat{k}+i\hat{p} \); \( \hat{u} \)
is then a compact real form of \( \hat{g}_{\mathbf{C}} \). (\emph{E.g.}, for
\( \hat{g}=\hat{s}\hat{l}\left( n,\mathbf{R}\right)  \) the Cartan decomposition
is the decomposition of a matrix into its antisymmetric and symmetric part;
in this case \( \theta X=-^{\mathrm{tr}}X \), \( \hat{k}=\hat{s}\hat{o}\left( n\right)  \),
\( \hat{p}=\left\{ \textrm{symmetric matrices}\right\}  \) and \( \hat{k}+i\hat{p}=\hat{s}\hat{u}\left( n\right)  \)).
If \( \hat{g}_{\mathbf{C}} \) is a semisimple Lie algebra over \( \mathbf{C} \),
\( \hat{u} \) its compact real form, then \( \hat{g}_{\mathbf{C}}=\hat{u}+i\hat{u} \)
is a Cartan decomposition of \( \hat{g}_{\mathbf{C}} \). (\emph{E.g.}, for
\( \hat{g}_{\mathbf{C}}=\hat{s}\hat{l}\left( n,\mathbf{C}\right)  \) it is
the decomposition of a matrix into its skew-hermitian and hermitian part). 

Since
\begin{equation}
\label{unomenouno}
\theta =1\; \textrm{on}\; \hat{k}\qquad \textrm{and}\qquad \theta =-1\: \textrm{on}\: \hat{p}
\end{equation}
then, for all \(X,Y\in\hat{g}\), 
\begin{eqnarray}
\theta \, \mathrm{ad}\left( X\right) Y & = & \mathrm{ad}\left( \theta X\right) \theta Y\label{thetaad} \\
\theta \, \mathrm{Ad}\left( \exp \left( X\right) \right) Y & = & \mathrm{Ad}\left( \exp \left( \theta X\right) \right) \theta Y\label{thetaAd} 
\end{eqnarray}
where \( \mathrm{Ad} \) is the adjoint representation of \( G \) on \( \hat{g} \)
(\emph{e.g.} \( \mathrm{Ad} \)\( \left( g\right) X=gXg^{-1} \) if \( G \),
\( \hat{g} \) are linear), so that \( \mathrm{Ad}\left( \exp \left( X\right) \right) =\exp \left( \mathrm{ad}\left( X\right) \right)  \).

Let \( \hat{h} \) be a maximal Abelian subalgebra of \( \hat{g} \), 
with \( \hat{a}=\hat{h}\cap \hat{p} \) as large as possible,
and let \( \hat{a}_{1}=\hat{h}\cap \hat{k} \). Then \( \hat{h}_{\mathbf{R}}=\hat{a}+i\hat{a}_{1} \)
and \( \hat{h}_{\mathbf{C}}=\hat{h}_{\mathbf{R}}+i\hat{h}_{\mathbf{R}} \) is
a Cartan subalgebra of \( \hat{g}_{\mathbf{C}} \). (\emph{E.g.}, for \( \hat{g}=\hat{s}\hat{l}\left( n,\mathbf{R}\right)  \),
\( \hat{a}_{1}=\left\{ 0\right\}  \) and \( \hat{a}=\hat{h}_{\mathbf{R}} \)
are the real diagonal matrices with null trace). Let \( P \) be the set of
positive roots that do not vanish identically on \( \hat{a} \) (for \( \hat{s}\hat{l}\left( n,\mathbf{R}\right)  \)
this just coincides with the set of positive roots). Let 
\[
\hat{n}_{\mathbf{C}}=\sum _{\alpha \in P}\hat{g}_{\mathbf{C}}^{\alpha },\qquad \hat{n}=\hat{n}_{\mathbf{C}}\cap \hat{g}\]
 Then \( \hat{n}_{\mathbf{C}} \) and \( \hat{n} \) are nilpotent Lie algebras
and one has the direct vector space sum 
\begin{equation}
\label{iwasawa}
\hat{g}=\hat{k}+\hat{a}+\hat{n}
\end{equation}
 (Iwasawa decomposition). From \emph{ii)} it follows that \( \hat{a} \) is
orthogonal with respect to \( B \) to both \( \hat{k} \) and \( \hat{n} \),
and that \( \hat{n} \) is orthogonal to itself. The decomposition (\ref{iwasawa})
exponentiates to 
\begin{equation}
\label{iwagroup}
G=KAN
\end{equation}
 where \( K=\exp \, \hat{k} \), \( A=\exp \, \hat{a} \), \( N=\exp \, \hat{n} \),
\emph{i.e.}, the mapping 
\begin{equation}
\label{diffeo}
\left( k,a,n\right) \rightarrow kan
\end{equation}
 is a diffeomorphism of \( K\times A\times N \) onto \( G \). (\emph{E.g.},
for \( \hat{s}\hat{l}\left( n,\mathbf{R}\right)  \), \( \hat{k} \) is the
rotation algebra, \( \hat{a} \) is the algebra of diagonal matrices with null
trace, \( \hat{n} \) is the algebra of strictly triangular matrices; at the
group level this decomposition can be naturally obtained through the Gram-Schmidt
orthogonalization process).

To the root decomposition (\ref{rootdec}) of \( \hat{g}_{\mathbf{C}} \) there
corresponds a restricted root decomposition 
\[
\hat{g}=\hat{a}+\hat{m}+\sum _{\alpha '\in \Delta '}\hat{g}^{\alpha '}\]
 where \( \hat{m} \) are the elements of \( \hat{k} \) which commute with
all the elements of \( \hat{a} \), \( \Delta ' \) is the set of nonzero restrictions
\( \alpha ' \) of \( \alpha  \) to \( \hat{a} \) and 
\[
\hat{g}^{\alpha '}=\left\{ X\in \hat{g}\, :\, \mathrm{ad}(H)\, X=\alpha '\left( H\right) X\, \textrm{for all }H\in \hat{a}\right\} \]
 Since different roots \( \alpha  \) can have the same restriction \( \alpha ' \)
to \( \hat{a} \), the multiplicity 
\begin{equation}
\label{multiplicities}
m_{\alpha '}=\dim \hat{g}^{\alpha '}
\end{equation}
 can be higher than 1 for some algebras. The Cartan involution changes the sign
of the restricted roots: 
\begin{equation}
\label{cartancambia}
\theta \hat{g}^{\alpha '}=\hat{g}^{-\alpha '}
\end{equation}
 The algebra \( \hat{k} \) can therefore be decomposed as 
\begin{equation}
\label{kesp}
\hat{k}=\hat{m}+\sum _{\alpha '\in P'}\left( \hat{g}^{\alpha '}+\theta\,\hat{g}^{\alpha '}\right) 
\end{equation}
 A Weyl basis \( \left\{ X_{\alpha }\right\}  \) for \( \hat{g}_{\mathbf{C}} \)
can be projected on \( \hat{g} \) by 
\begin{equation}
\label{base}
Y_{\alpha }=\frac{1}{2}\left( X_{\alpha }+\sigma X_{\alpha }\right) 
\end{equation}
 so that \( \hat{g}^{\alpha '}=\sum \mathbf{R}Y_{\alpha } \) where the sum
runs over all \( \alpha \in \Delta  \) whose restriction to \( \hat{a} \)
is \( \alpha ' \).

On a Lie group one can define the (right invariant) Haar measure 
\[
\mathrm{d}g=\bigwedge _{l=1}^{\dim G}g^{-1}\, \delta _{l}g\]
 where \( \delta _{l}g=g\, \delta _{l}X \) and the \( \delta _{l}X \) are
a basis of 1-forms on \( \hat{g} \). Semisimple Lie groups are unimodular,
\emph{i.e.} such that \( \left| \det \mathrm{Ad}\left( \mathrm{g}\right) \right| =1 \)
for each \( g\in G \), so that right invariant measures are also left invariant.
The Iwasawa decomposition induces the decomposition 
\begin{equation}
\label{decmes}
\mathrm{d}g=j\left( k,a,n\right) \mathrm{d}k\, \mathrm{d}a\, \mathrm{d}n
\end{equation}
 where \( j \) is a Jacobian factor. The groups \( G \), \( K \), \( A \),
\( N \) are all unimodular, therefore both sides of (\ref{decmes}) are invariant
under the right and left translations \( g\rightarrow k_{\mathrm{R}}g \), \( g\rightarrow gn_{\mathrm{L}} \)
with \( k_{\mathrm{R}}\in K \) and \( n_{\mathrm{L}}\in N \), so that \( j\left( k,a,n\right) =j\left( a\right)  \).
On the other hand, the l.h.s of (\ref{decmes}) is also invariant under \( g\rightarrow ga^{-1} \),
which corresponds to \( kan\rightarrow ke\cdot \left( ana^{-1}\right)  \) (with
\( e \) the identity element), so that 
\begin{eqnarray*}
j\left( a\right)  & = & j\left( e\right) \, \frac{\partial \left( ana^{-1}\right) }{\partial \left( n\right) }\\
 & = & j\left( e\right) \, \det \left( \left. \mathrm{Ad}\left( a\right) \right| _{\hat{n}}\right) \\
 & = & j\left( e\right) \, \exp \left[ \mathrm{Tr}\, \left. \mathrm{ad}\left( \exp ^{-1}a\right) \right| _{\hat{n}}\right] \\
 & = & j\left( e\right) \, \exp \left[ \sum _{\alpha \in P}\alpha \left( \exp ^{-1}a\right) \right] 
\end{eqnarray*}
 If the measures on \( G \), \( K \), \( A \), \( N \) are conveniently
normalized we can assume that \( j\left( e\right) =1 \), so 
\begin{equation}
\label{iwwames}
\mathrm{d}g=\exp \left[ 2\rho \left( \exp ^{-1}a\right) \right] \, \mathrm{d}k\, \mathrm{d}a\, \mathrm{d}n
\end{equation}
 with 
\begin{equation}
\label{semisomma}
\rho =\frac{1}{2}\sum _{\alpha \in P}\alpha 
\end{equation}

Let \( \tau  \) be a skew-hermitian representation of the compact form \( \hat{u} \)
on a finite dimensional Hilbert space \( V \). By analytically extending \( \tau  \)
to the complexification \( \hat{g}_{\mathbf{C}}=\hat{u}+i\hat{u} \) and then
restricting to the real form \( \hat{g}=\hat{k}+\hat{p} \) one obtains a representation
of the real algebra \( \hat{g} \) (Weyl unitary trick). If \( K\in \hat{k} \),
\( P\in \hat{p} \), then \( \tau \left( K\right)  \) and \( \tau \left( iP\right)  \)
are skew-hermitian, while \( \tau \left( P\right)  \) is hermitian. In other
words, 
\[
\tau ^{\dagger }\left( X\right) =-\tau \left( \theta X\right) \]
 Denoting \( T\left( \exp \left( X\right) \right) =\exp \left( \tau \left( X\right) \right)  \),
this property exponentiates \cite[p. 144]{knapp86} to 
\begin{equation}
\label{ultimocartan}
T^{\dagger }\left( g\right) =T^{-1}\left( \Theta g\right) 
\end{equation}


\begin{thebibliography}{10}
\bibitem{ak87}L.~Arnold and W.~Kliemann. Large deviations of linear stochastic differential
equations. \newblock In H.J. Engelbert and W.~Schmidt, editors, \emph{Stochastic
Differential Systems. Proceedings}, volume~96 of \emph{Lect. Notes in Control
and Information Sciences}, pages 117--151, Berlin, 1987. Springer. \newblock Eisenach
1986.
\bibitem{aw86}For a review of results, see: L.~Arnold and V.~Wihstutz, editors. \newblock {\emph{Lyapunov Exponents. Proceedings}},
volume 1186 of \emph{Lect. Notes in Mathematics}, Berlin, 1986. Springer. \newblock Bremen
1984.
\bibitem{beenakker97}For a review of results, see: C.W.J. Beenakker. \newblock Random-matrix theory
of quantum transport. \newblock {\emph{Rev. Mod. Phys.}}, \textbf{69}:731, 1997.
\bibitem{br93a}C.W.J. Beenakker and B.~Rejaei. \newblock Nonlogarithmic repulsion of transmission
eigenvalues in a disordered wire. \newblock {\emph{Phys. Rev. Lett.}}, \textbf{71}:3689--3692,
1993.
\bibitem{benettin84}G.~Benettin. \newblock Power law behaviour of Lyapunov exponents in some conservative
dynamical systems. \newblock {\emph{Physica D}}, \textbf{13}:211--213, 1984.
\bibitem{bggs80}G.~Benettin, L.~Galgani, A.~Giorgilli, and J.M. Strelcyn. \newblock Lyapunov
characteristic exponents for smooth dynamical systems and for Hamiltonian systems:
a method for computing all of them. \newblock {\emph{Meccanica}}, \textbf{15}:9,
1980.
\bibitem{bgk98}D.~Bernard, K.~Gaw\c{e}dzki, and A.~Kupiainen. \newblock Slow modes in passive
advection. \newblock {\emph{J. Stat. Phys.}}, \textbf{90}:519, 1998.
\bibitem{cgk94a}A.S.~Cattaneo, A.~Gamba, and I.~Kolokolov. \newblock Statistics of the one-electron
current in a one-dimensional mesoscopic ring at arbitrary magnetic fields. \newblock {\emph{J. Stat. Phys.}},
\textbf{76}:1065--1074, 1994.
\bibitem{caselle95}M.~Caselle. \newblock Distribution of transmission eigenvalues in disordered
wires. \newblock {\emph{Phys. Rev. Lett.}}, \textbf{74}:2776--2779, 1995.
\bibitem{cfk98}M.~Chertkov, G.~Falkovich, and I.~Kolokolov. \newblock Intermittent dissipation
of passive scalar in turbulence. \newblock {\emph{Phys. Rev. Lett.}}, \textbf{80}:2121--2124,
1998.
\bibitem{cfkl95a}M.~Chertkov, G.~Falkovich, I.~Kolokolov, and V.~Lebedev. \newblock Statistics
of a passive scalar advected by a large scale two-dimensional velocity field:
analytic solution. \newblock {\emph{Phys. Rev. E}}, \textbf{51}:5609--5627,
1995.
\bibitem{cfkv99}M.~Chertkov, G.~Falkovich, I.~Kolokolov, and M.~Vergassola. \newblock Small-scale
turbulent dynamo. \newblock {\emph{Phys. Rev. Lett.}}, \textbf{83}:4065, 1999.
\bibitem{cgk94}M.~Chertkov, A.~Gamba, and I.~Kolokolov. \newblock Exact field-theoretical description
of passive scalar convection in an \( n \)-dimensional long range velocity
field. \newblock {\emph{Phys. Lett. A}}, \textbf{192}:435--443, 1994.
\bibitem{cpv93}For a review of results, see: A.~Crisanti, G.~Paladin, and A.~Vulpiani. \newblock {\emph{Products of random matrices in statistical physics}}.
\newblock Springer-Verlag, Berlin-Heidelberg, 1993.
\bibitem{dorokhov82}O.~N. Dorokhov. \newblock Transmission coefficient and the localization length
of an electron in \( {N} \) bound disordered chains. \newblock {\emph{JETP Lett.}},
\textbf{36}:318--321, 1982.
\bibitem{eg00}J.-P. Eckmann and O.~Gat. \newblock Hydrodynamic Lyapunov modes in translation
invariant systems. \newblock {\emph{J. Stat. Phys.}}, \textbf{98}:775--798,
2000.
\bibitem{fgv01}For a review of results, see: G.~Falkovich, K.~Gaw\c{e}dzki, and M.~Vergassola.
\newblock Particles and fields in fluid turbulence. \newblock {\emph{Rev. Mod. Phys.}},
\textbf{73}:913--975, 2001.
\bibitem{fkl98}G.~Falkovich, V.~Kazakov, and V.~Lebedev. \newblock Particle dispersion in a
multidimensional random flow with arbitrary temporal correlations. \newblock {\emph{Physica A}},
\textbf{249}:36--46, 1998.
\bibitem{fh65}R.~P. Feynman and A.~R. Hibbs. \newblock {\emph{Quantum mechanics and path integrals}}.
\newblock Mc-Graw Hill, New York, 1965.
\bibitem{gk96}A.~Gamba and I.~Kolokolov. \newblock The Lyapunov spectrum of a continuous product
of random matrices. \newblock {\emph{J. Stat. Phys.}}, \textbf{85}:489--499,
1996.
\bibitem{gk99}A.~Gamba and I.~Kolokolov. \newblock Dissipation statistics of a passive scalar
in a multidimensional smooth flow. \newblock {\emph{J. Stat. Phys.}}, \textbf{94}:759--777,
1999.
\bibitem{gy60}I.~M. Gel'fand and A.~M. Yaglom. \newblock Integration in function spaces and
its applications in quantum physics. \newblock {\emph{J. Math. Phys.}}, \textbf{1}:48,
1960.
\bibitem{helgason78}S.~Helgason. \newblock {\emph{Differential geometry, {Lie} groups, and symmetric spaces}}.
\newblock Academic Press, 1978.
\bibitem{helgason84}S.~Helgason. \newblock {\emph{Groups and geometric analysis}}. \newblock Academic
Press, 1984.
\bibitem{huffman90}A.~H\"{u}ffman. \newblock Disordered wires from a geometric viewpoint. \newblock {\emph{J. Phys. A}},
\textbf{23}:5733--5744, 1990.
\bibitem{iwasawa49}K.~Iwasawa. \newblock On some types of topological groups. \newblock {\emph{Ann. of Math.}},
\textbf{50}:507--558, 1949.
\bibitem{knapp86}A.W. Knapp. \newblock {\emph{Representation theory of semisimple groups}}. \newblock Princeton
Un. Press, 1986.
\bibitem{knapp96}A.W. Knapp. \newblock {\emph{Lie groups beyond an introduction}}. \newblock Birkhäuser,
1996.
\bibitem{kolokolov86}I.V. Kolokolov. \newblock Functional representation for the partition function
of the quantum Heisenberg ferromagnet. \newblock {\emph{Phys. Lett. A}}, \textbf{114}:99--104,
1986.
\bibitem{kolokolov90}I.V. Kolokolov. \newblock Functional integration for quantum magnets: new method
and new results. \newblock {\emph{Ann. of Phys.}}, \textbf{202}:165--185, 1990.
\bibitem{kolokolov93}I.V. Kolokolov. \newblock The method of functional integration for one-dimensional
localization, higher correlators, and the average current flowing in a mesoscopic
ring in an arbitrary magnetic field. \newblock {\emph{JETP}}, \textbf{76}:1099,
1993.
\bibitem{oseledec68}V.I. Oseledec. \newblock The multiplicative ergodic theorem. The Lyapunov characteristic
numbers of dynamical systems. \newblock {\emph{Trans. Moscow Math. Soc.}}, \textbf{19}:197--231,
1968.
\bibitem{pv86}G.~Paladin and A.~Vulpiani. \newblock Scaling law and asymptotic distribution
of Lyapunov exponents in conservative dynamical systems with many degrees of
freedom. \newblock {\emph{J. Phys. A}}, \textbf{19}:1881, 1986.
\bibitem{zinn96}J.~Zinn-Justin. \newblock {\emph{Quantum field theory and critical phenomena}}.
\newblock Clarendon Press, Oxford, 1996.
\end{thebibliography}
\end{document}